\documentclass[12pt,preprint]{aastex}
\def\hal{H$\alpha$}
\def\hbeta{H$\beta$}
\def\be{\begin{equation}}
\def\ee{\end{equation}}

\def\m{~$\mu$m}

\def\NII  {[\ion{N}{2}]}
\def\NII  {[\ion{N}{2}]}

\def\OII  {[\ion{O}{2}]}
\def\OIII {[\ion{O}{3}]}

\def\Nsourcesa{98}
\def\Nsourcesb{208}
\def\sqdeg{2}
\def\rhoa{0.009}
\def\rhob{0.014}

\begin {document}
%\slugcomment{\scriptsize \today \hskip 0.2in Version 1.0}

\title{The Wyoming Survey for H$\alpha$.  I.  Initial Results at $z\sim0.16$ and $0.24$}
\shorttitle{The Wyoming Survey for H$\alpha$}

\author{Daniel~A.~Dale, Rebecca~J.~Barlow, Seth~A.~Cohen, L.~Clifton~Johnson, ShiAnne~M.~Kattner, Christine~A.~Lamanna, Carolynn~A.~Moore, Micah~D.~Schuster, Jacob~W.~Thatcher}
\affil{Department of Physics and Astronomy, University of Wyoming, Laramie, WY 82071}
%\affil{ddale@uwyo.edu}

\begin {abstract}
The {\it Wyoming Survey for H$\alpha$}, or WySH, is a large-area, ground-based, narrowband imaging survey for \hal-emitting galaxies over the latter half of the age of the Universe.  The survey spans several square degrees in a set of fields of low Galactic cirrus emission.  The observing program focuses on multiple $\Delta z \sim 0.02$ epochs from $z\sim0.16$ to $z\sim0.81$ down to a uniform (continuum+line) luminosity at each epoch of $\sim10^{33}$~W uncorrected for extinction (3$\sigma$ for a 3\arcsec\ diameter aperture).  First results are presented here for \Nsourcesa +\Nsourcesb\ galaxies observed over approximately \sqdeg\ square degrees at redshifts $z\sim0.16$ and 0.24, including preliminary luminosity functions at these two epochs.  These data clearly show an evolution with lookback time in the volume-averaged cosmic star formation rate.  Integrals of Schechter fits to the extinction-corrected \hal\ luminosity functions indicate star formation rates per co-moving volume of \rhoa\ and \rhob~$h_{70}~M_\odot$~yr$^{-1}$~Mpc$^{-3}$ at $z \sim 0.16$ and 0.24, respectively.  The formal uncertainties in the Schechter fits, based on this initial subset of the survey, correspond to uncertainties in the cosmic star formation rate density at the $\gtrsim$40\% level; the tentative uncertainty due to cosmic variance is 25\%, estimated from separately carrying out the analysis on data from the first two fields with substantial datasets.
\end {abstract}

\keywords{galaxies: evolution --- galaxies: luminosity function}
 
\section {Introduction}
How do galaxies evolve?  Such a broad query covers a wide variety of current research interests.  One specific approach to answering this question lies in tracking the volume density of the cosmic star formation rate as a function of redshift.  Early pioneering efforts by Lilly et al. (1996), Madau et al. (1996), and Steidel et al. (1996) forged a scenario in which the population of star-forming galaxies evolves out to a redshift of approximately unity, and beyond this redshift the star formation rate density plateaus or slowly declines.  This general picture has since been confirmed through various imaging and spectroscopy surveys carried out at ultraviolet, optical, infrared, and submillimeter wavelengths (e.g., Blain et al. 1999; Rowan-Robinson 2001; Tresse et al. 2002; Baldry et al. 2002; Wilson et al. 2002; Mann et al. 2002; Glazebrook et al. 2003; Hippelein et al. 2003, Heavens et al. 2004, Gabasch et al. 2004, P\'erez-Gonz\'alez et al. 2005, Le Floc'h et al. 2005, Thompson et al. 2006, Babbedge et al. 2006, Sawicki \& Thompson 2006, Ly et al. 2007; Shioya et al. 2007; Westra \& Jones 2007).  The rate of this drop-off provides an important constraint to cosmologists--understanding the physical processes involving stellar birth, life, and death, and how these processes feed back into the interstellar medium is crucial to our overall understanding of galaxy formation and evolution (e.g. Somerville, Primack \& Faber 2001; Ascasibar et al. 2002).  Galaxy merger rates, the frequency of starburst galaxies compared to active galactic nuclei, the production of heavy metals and dust, and the origin of the cosmic background levels (Dole et al. 2006) are also closely linked to the cosmic star formation rate. 

While the decrease in the cosmic star formation rate density since $z\sim1$ has been empirically parametrized as $(1+z)^\gamma$, the strength of this evolution is not well determined.  Values for the exponent range from $\gamma\approx1.5$ (e.g. Cowie, Songaila \& Barger 1999; Wilson et al. 2002) to $\gamma\approx4$ (e.g. Lilly et al. 1996; Blain et al. 1999; Tresse et al. 2002, P\'erez-Gonz\'alez et al. 2005, Le Floc'h et al. 2005, Babbedge et al. 2006), a discrepancy of a factor of six in the change in the star formation rate from $z=0$ to $z=1$.  Constraining galaxy evolution through measuring changes in the star formation rate density at even higher redshifts is a viable and important goal.  However, a deep understanding of galaxy evolution, from the era of galaxy formation until now, requires an accurate foundation that is based on observations of the most recent epochs.
 
The current conflict over the strength of the evolutionary trend can be addressed by carrying out a uniformly sensitive survey for star formation in hundreds of
galaxies at each of several epochs.  The {\it Wyoming Survey for H$\alpha$}, or WySH, is a large-area, uniform survey for star-forming galaxies at several different epochs that span the latter half of the age of the 
Universe\footnote{Results from the $z\sim0$ \hal\ surveys of Gallego et al. (1995), Gronwall et al. (1999), and Hanish et al. (2006) will be used for the star formation rate density at the current epoch to supplement our work.}.  The specific aim of the survey is to utilize narrowband imaging of \hal-emitting galaxies to conclusively determine the strength of the evolution in the cosmic star formation rate density since $z\sim0.8$, to provide a firm, low redshift baseline for studies of star-forming sources at higher redshift.  The purpose of this paper is to introduce the survey, in particular the optical imaging program, and to present initial results from the two most recent epochs studied.  Though only preliminary results are presented in this contribution, the survey will ultimately cover volumes at each epoch large enough to minimize cosmic variance due to large scale structure and to robustly quantify any dependencies on environment. 

In this paper we outline the optical portion of the survey in \S~\ref{sec:experiment}, present initial results from $z\sim0.16$ and $z\sim0.24$ in \S~\ref{sec:results}, and summarize in \S~\ref{sec:summary}.  The cosmology assumed is $H_0=70~h_{70}$~km~s$^{-1}$~Mpc$^{-1}$, $\Omega_{\rm m}=0.3$, and $\Omega_\lambda=0.7$.  

\section{The Survey}
\label{sec:experiment}

The optical portion of the survey is being carried out via large blocks of dark time on the Wyoming Infrared Observatory 2.3~m telescope (WIRO)\footnote{The near-infrared portion of the survey (z$\sim$0.81) is being completed as a collaborative effort with J. Lee, R. Finn, and D. Eisenstein using the Bok 2.3~m and Mayall 4~m telescopes on Kitt Peak.  Results from that effort will be presented elsewhere.}.  Unlike many previous efforts which piece together the star formation history using disparate surveys and techniques/tracers at different epochs, the data are being uniformly obtained over the same fields and to the same sensitivity to luminosity over the different epochs.  The survey is being carried out over 3.5-4 square degrees from multiple ``blank'' fields selected to minimize foreground contamination (see \S~\ref{sec:field_selection}).  By design, these fields are additionally pre-selected to overlap with the target areas of deep infrared and ultraviolet surveys.  Not only will this survey provide a statistically robust measure of the evolving star formation history of the Universe, it will also enable important parallel science to be pursued.  For example, combining \hal, ultraviolet, and infrared data will help us to understand the extent to which the standard individual star formation rate indicators are biased (Wilson et al. 2002), and to study the evolution of ultraviolet and optical extinction by interstellar dust over time (e.g., Reddy et al. 2006; Burgarella et al. 2007; Iglesias-P\'aramo et al. 2007; Moore et al., in preparation).  In addition, a large volume, multi-epoch survey will sample well the cluster, group, and field environments, leading to strong constraints on the environmental threshold that provides the balance between star formation inducement and truncation that occurs in the vicinity of clusters (e.g., Dale et al. 2001; Balogh et al. 2002; Finn et al. 2005).

\subsection{H$\alpha$ as a Star Formation Tracer}
Astronomers rely on a wide range of observational approaches for quantifying star formation.  The ultraviolet, infrared, and radio restframe continua all provide useful measures of the star formation rate, as do line fluxes in the ultraviolet, optical, infrared, and sub/millimeter (Kennicutt 1998; Calzetti et al. 2007; Kennicutt et al. 2007).  However, for local Universe systems \hal\ photons, indicative of ionized gaseous regions and hence directly related to the massive stellar population, provide the conventional standard by which to gauge star formation (Kennicutt 1998).  \hal\ is typically the intrinsically strongest optical emission line and it suffers less extinction than other traditional indicators that lie at shorter wavelengths.  Moreover, \hal\ fluxes are technically much simpler to obtain than star formation measures at longer wavelengths.  Thus a feasible and direct route to measuring galaxy evolution, one that is well-calibrated by local Universe measurements and is accessible from a 2~m class ground-based telescope equipped with a wide-field camera, would be to survey the Universe in \hal\ at several different epochs.

\subsection{Field Selection}
\label{sec:field_selection}
In addition to carrying out \hal\ observations over multiple epochs, we are observing multiple fields (Table~\ref{tab:fields}; Figures~\ref{fig:wysh1}-\ref{fig:elaisn1}).  Three separate fields help to dampen the effects of cosmic variance (Davis \& Huchra 1982; Oliver et al. 2000; Huang et al. 2007).  Moreover, the spacing of the fields in right ascension and their high declinations enable year-round observations from Jelm Mountain, located at a terrestrial latitude of +41\degr\ and an elevation of 9656~feet (2943~m) above sea level, thus maximizing the number of usable nights.  These fields have been chosen to minimize foreground contamination by zodiacal and Galactic dust, bright stars, and nearby galaxies.  Our pointings in the ELAIS-N1 and Lockman Hole fields overlap with deep surveys that have been carried out in the infrared and ultraviolet.  The SWIRE Spitzer Legacy project provides maps of these regions at 3.6, 4.5, 5.8, 8.0, 24, 70, and 160\m\ (Lonsdale et al. 2003), while the GALEX Deep Imaging Survey offers 30~ks integrations in these regions at 1528 and 2271\AA, integrations that are much deeper than the canonical 0.1~ks integration per pointing of the GALEX all-sky survey (Martin et al. 2005).  Our third field, dubbed WySH~1, conveniently fills the right ascension gap between the Lockman Hole and ELAIS-N1, at a northern declination not covered by any other current deep field work.
 
\subsection{Observations and Data Processing}
\label{sec:observations}

\begin{figure}
 \plottwo{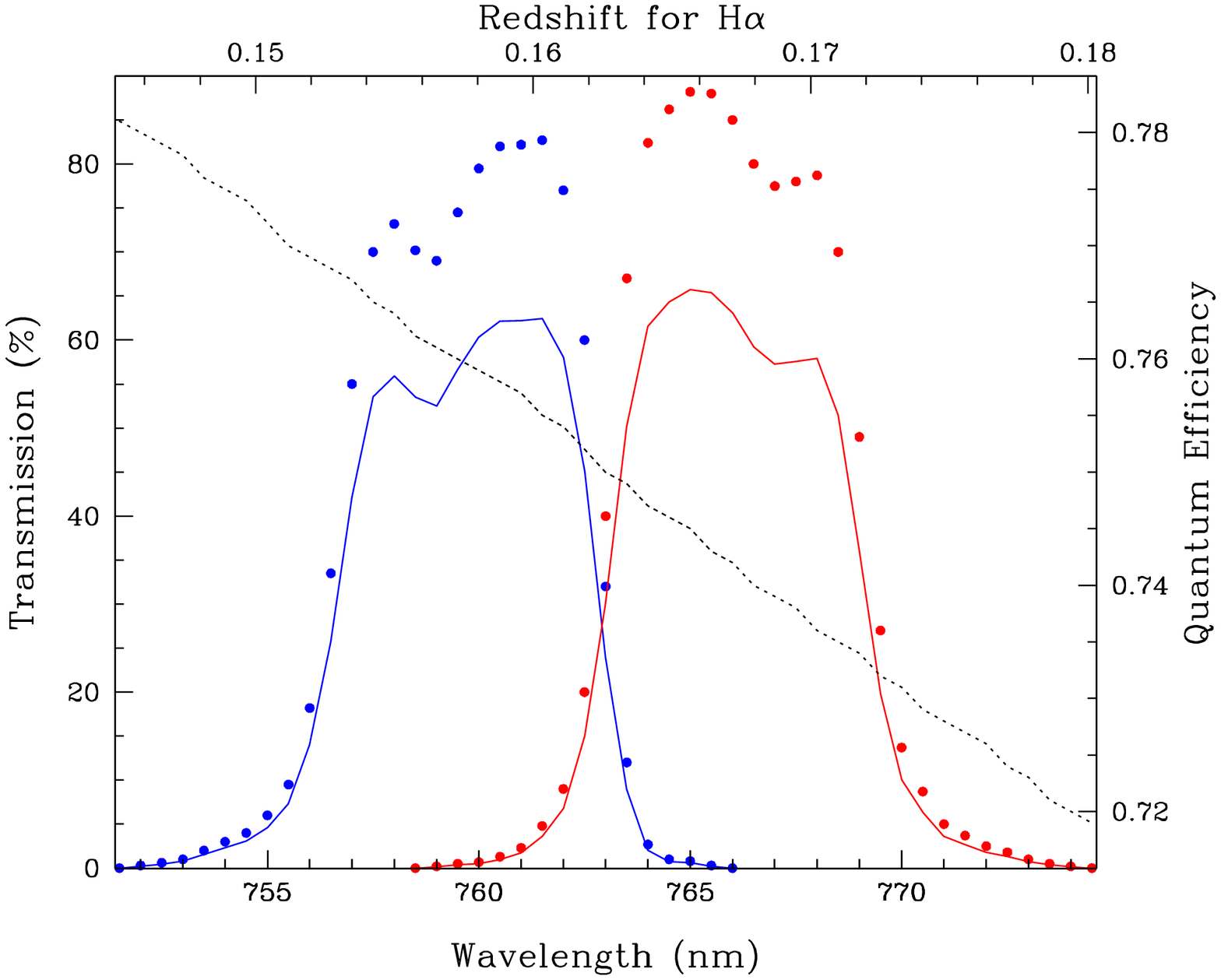}{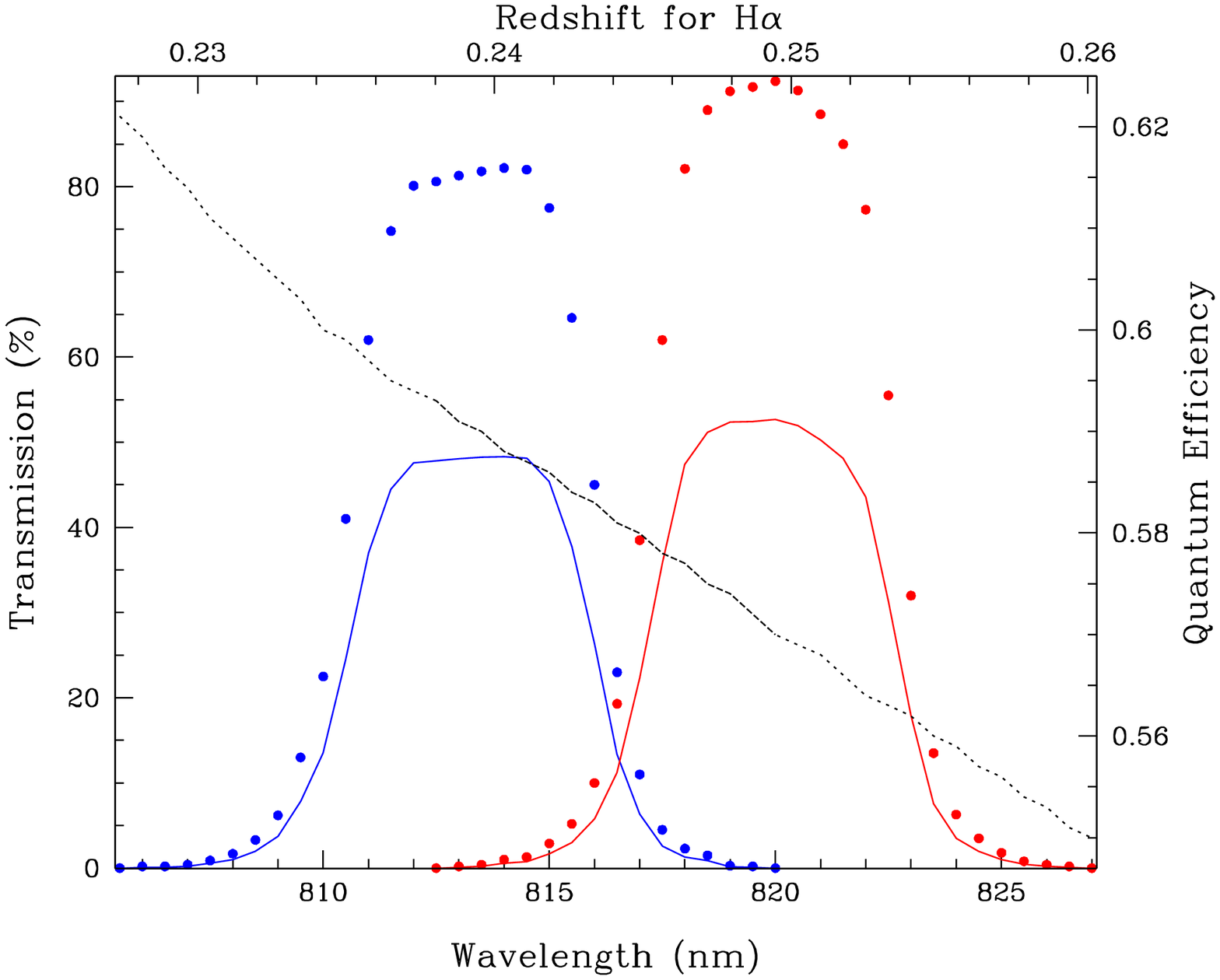}
 \caption{(a) The 7597$\AA$/7661$\AA$ filter pair.  (b)  The 8132$\AA$/8199$\AA$ filter pair.  The filter transmission profiles are plotted as filled circles, and the CCD quantum efficiency is shown as a dotted line (and enumerated along the righthand axis).  The combination of the two effects is shown with solid lines.} 
 \label{fig:filters1}
\end{figure}

\begin{figure}
 \plottwo{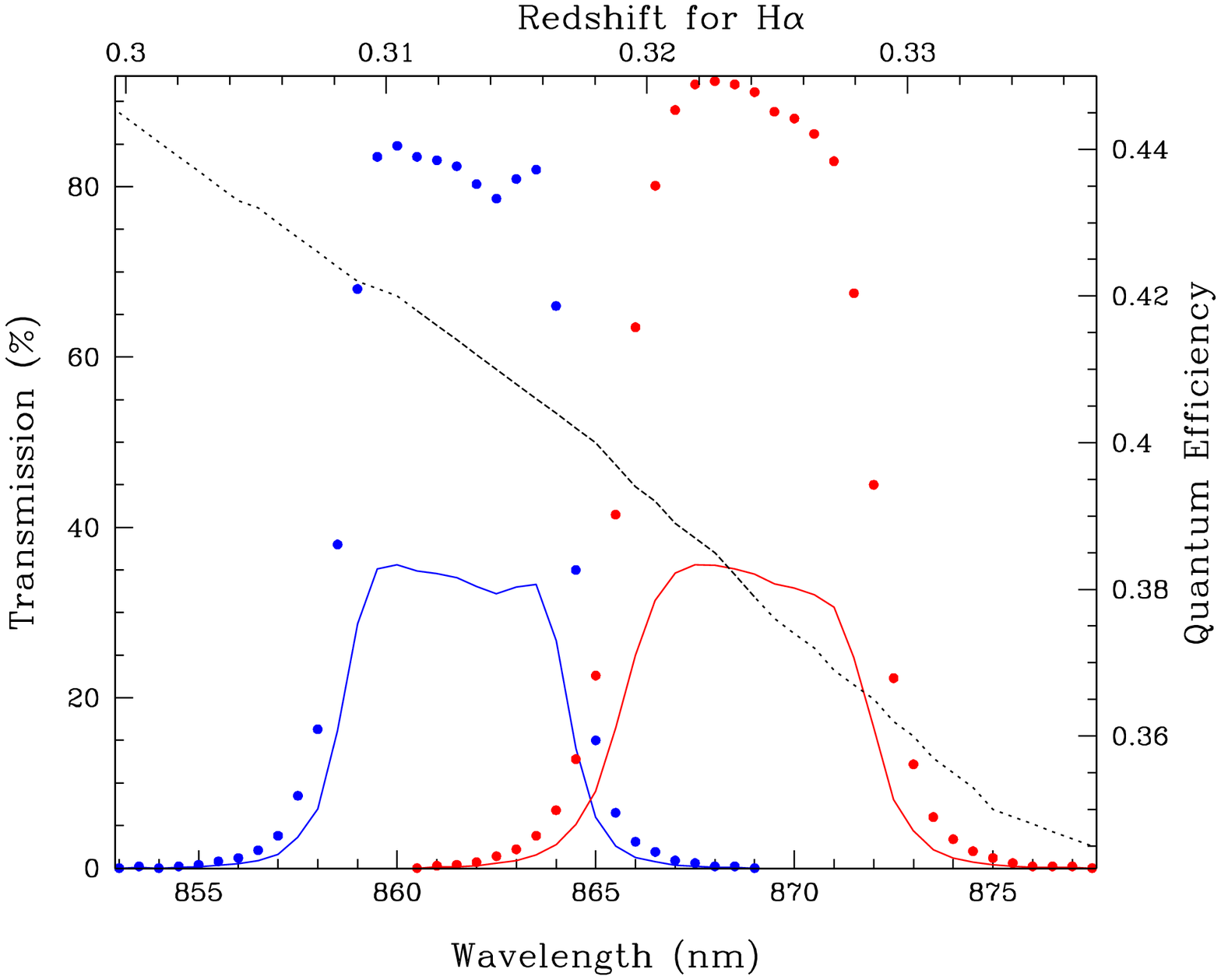}{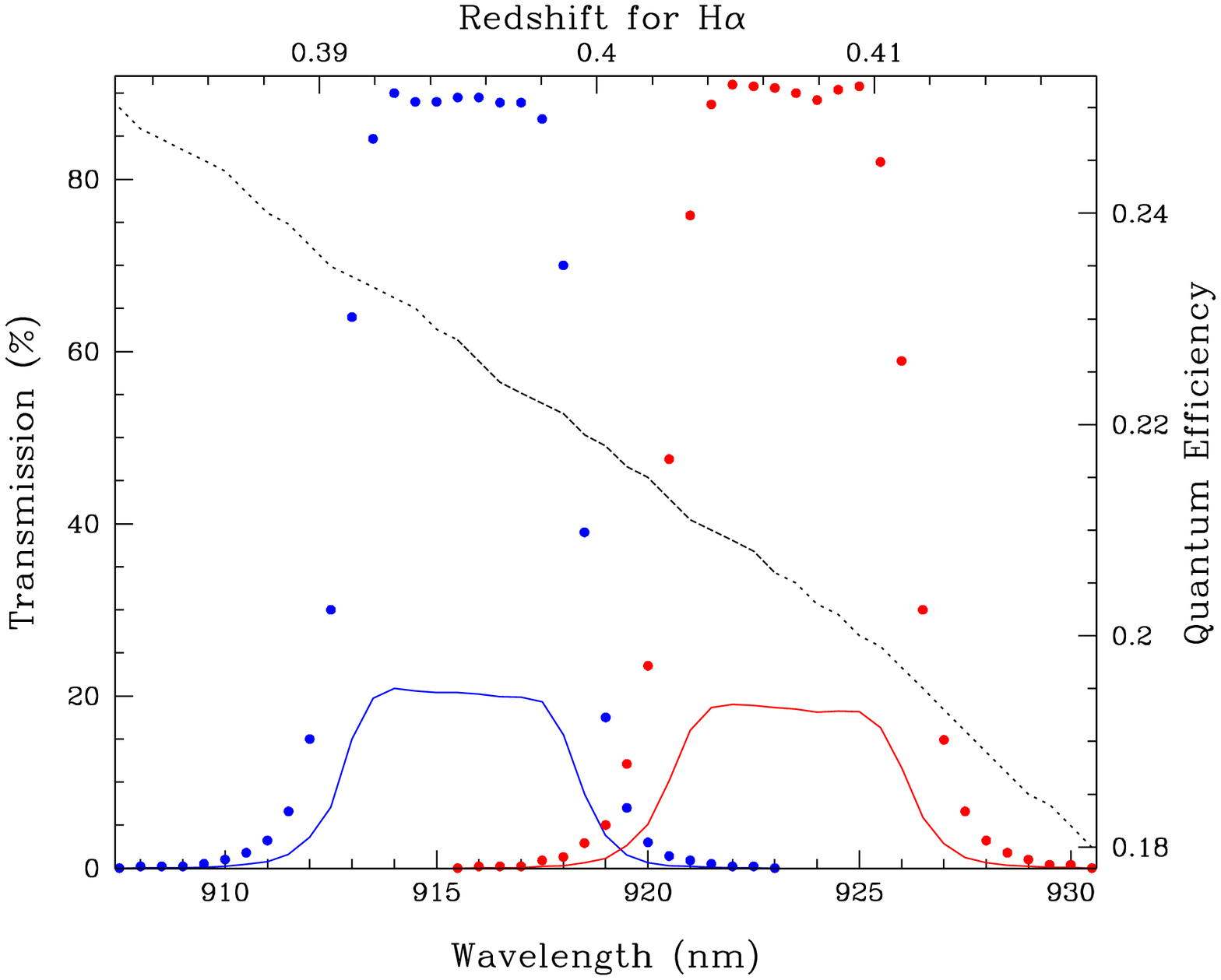}
 \caption{(a) The 8614$\AA$/8687$\AA$ filter pair.  (b)  The 9155$\AA$/9233$\AA$ filter pair.  The filter transmission profiles are plotted as filled circles, and the CCD quantum efficiency is shown as a dotted line (and enumerated along the righthand axis).  The combination of the two effects is shown with solid lines.} 
 \label{fig:filters2}
\end{figure}

Optical observations for the survey utilize the WIRO 2.3~m telescope.  WIRO is equipped with a 2048x2048 CCD camera at prime focus that has 0\farcs523 pixels and hence a 17\farcm9 field of view (Pierce \& Nations 2002).  While most narrowband imaging surveys rely on broadband imaging for the estimation of the continuum flux levels, we use wavelength-adjacent narrowband filters ($\sim60$\AA\ FWHM) to optimize the continuum subtraction.  Note that this approach leads to a more efficient observing program, since no time is ``wasted'' on obtaining broadband imaging.  The filter transmission profiles for the optical portion of this survey are provided in Figures~\ref{fig:filters1} \& \ref{fig:filters2}.  The filter widths and central wavelengths in Table~\ref{tab:filters} are computed via
\be
\Delta \equiv \int \mathcal{T}(\lambda) d\lambda
\ee
\be
\bar{\lambda} \equiv {\int \lambda \mathcal{T}(\lambda) d\lambda \over \int \mathcal{T}(\lambda) d\lambda}
\ee
where $\mathcal{T}$ reflects the combined transmittance of the filter and the quantum efficiency of the CCD, normalized to peak at unity (see Pascual, Gallego, \& Zamorano 2007).  Note that these values are for the typical observing temperature of 0\degr~C.

The basic unit of observation is a 300~s frame, though several are taken for each filter at each location; the individual 300~s frames are slightly dithered to improve image reconstruction.  IRAF/{\tt CCDPROC} is used to run the images through the standard preliminary data processing routines.  Bias levels and dark counts are removed, the images are trimmed to their usable regions (2048x2037 pixels), and the images are flat-fielded using twilight sky flats.  Fringing affects our images primarily at wavelengths of 8132\AA\ and longer, and all images suffer from non-uniform illumination that induce a mild halo effect if not removed.  The amplitude of fringing is at the 10\% level or smaller in terms of sky fluctuation, whereas the non-uniform illumination results in 1-2\% variations in the sky level across the field.  {\tt CCDPROC} is also used to correct for both of these effects, relying on the science frames themselves for fringe removal and on median skyflat images for the non-uniform illumination. 

The data presented here were obtained under generally good observing conditions during several observing runs spanning 2005, 2006, and 2007.  The seeing at Jelm Mountain is typically 1\farcs5 FWHM; the observed stellar PSFs from 300~s frames range from 1\farcs2-2\farcs0 for this survey.  Flux calibration is obtained via 30~s observations of spectrophotometric standards (Bohlin, Dickinson, \& Calzetti 2001).  Based on numerous repeated measurements of the same set of standard stars, the uncertainty in the flux calibration is 5\%.  In addition to frequently monitoring the night sky conditions by eye, the photometric quality for each frame is based on the relative fluxes in five foreground stars distributed across the field of view.  Ninety percent of the 300~s frames used in this presentation were obtained under photometric conditions (in which foreground stellar fluxes agree to within 2-3\% from frame to frame).  For the remaining 10\% of the frames taken in `nearly photometric' conditions, the median flux correction to account for thin cirrus cloud cover is a factor of 1.07.  Such a correction is applied to each image in conjunction with a correction to convert all fluxes to their equivalent values for zero airmass.

The multiple 300~s frames for a given field are aligned and stacked to create images with longer effective integrations (Table~\ref{tab:filters}).  The `stack' is an average frame constructed from IRAF/{\tt IMCOMBINE} with {\tt avsigclip} rejection and a weighting applied to each image that is based on the photometric quality described above.  Coordinate solutions are written to each stack header via IRAF/{\tt XYEQ} and IRAF/{\tt CCMAP}, based on the coordinate solutions in Digitized Sky Survey imaging and the pixel coordinates for several matched field stars in DSS and WySH frames.  Inspection of numerous sources spread across the fields of view of several stacks shows that WySH coordinates match DSS coordinates to within an arcsecond.

\subsection {Survey Sensitivity}
\label{sec:sensitivity}
The sensitivity of the survey is estimated in two ways, with the first based on the fluctuation of sky counts and the second based on the observed distribution of detections (the identification of sources is described in \S~\ref{sec:sextractor}).  In the first method, assuming a 3\arcsec\ diameter aperture and using the observed 3$\sigma$ sky fluctuations of $\sim 6.2 \cdot 10^{-20}$ and $\sim 3.7 \cdot 10^{-20}$~W~m$^{-2}$ at 7597/7661\AA\ and 8132/8199\AA, respectively, we find that the 3$\sigma$ limiting sensitivity to continuum/continuum+line luminosity in our survey is $\sim 5-7 \cdot 10^{32}$~W.  

\begin{figure}
 \plotone{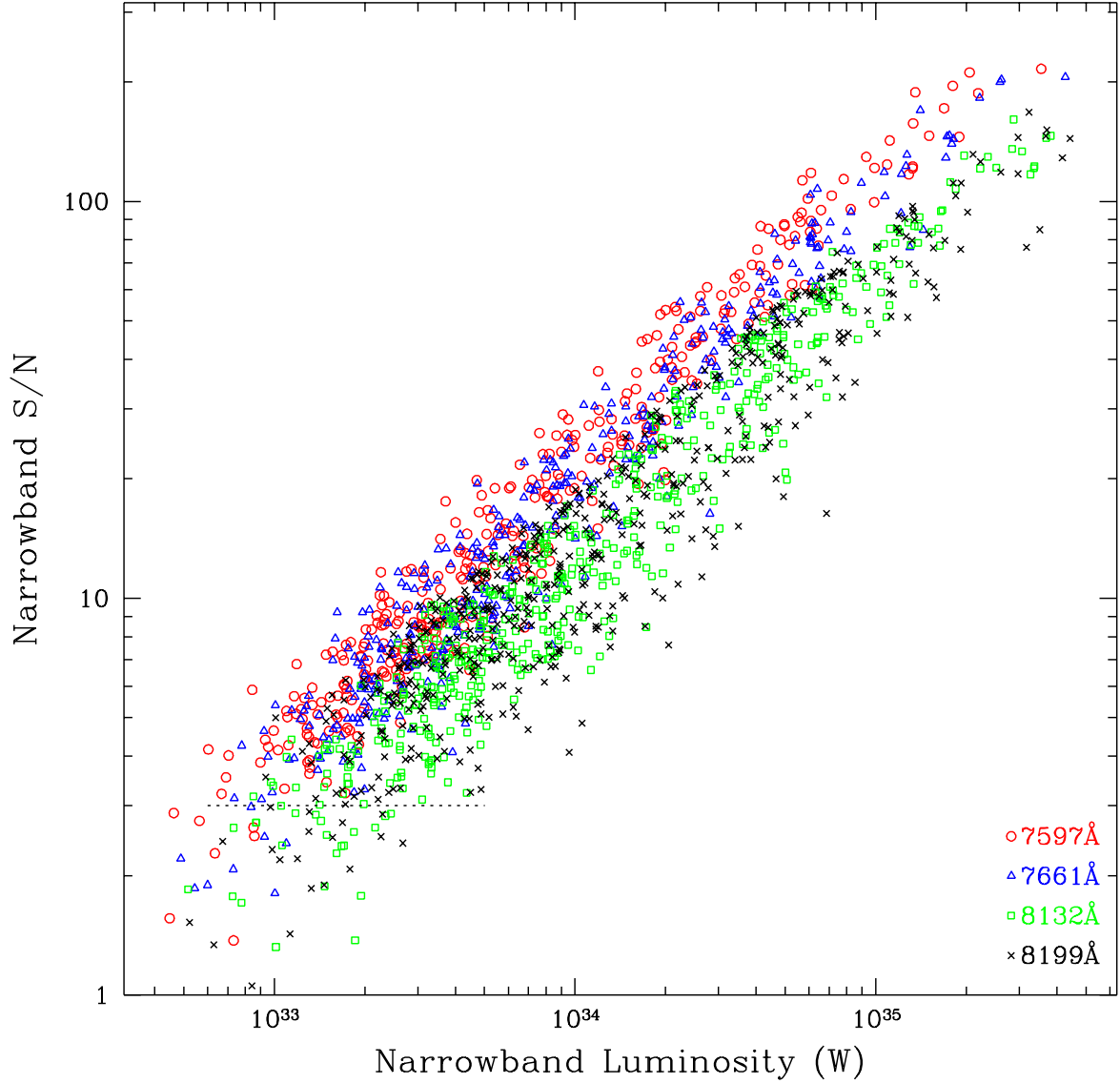}
 \caption{The distribution of signal-to-noise in the individual narrowband images, for sources at the targeted redshifts, as a function of narrowband luminosity.  A 3$\sigma$ cut corresponds to $\sim10^{33}$~W.}
 \label{fig:distribution}
\end{figure}

The noise characteristics within a fixed circular aperture does not necessarily capture the full sense of the survey sensitivity.  The seeing PSF is of course not constant from night to night, nor do we even utilize fixed aperture photometry for our source extraction (\S~\ref{sec:sextractor}).  Figure~\ref{fig:distribution} shows the relevant figure for the second method of estimating survey sensitivity.  In this figure we show the observed distribution of signal-to-noise and luminosity for all narrowband detections at the targeted redshifts.  As can be seen from the distribution, the 3$\sigma$ sensitivity to luminosity is $6 \cdot 10^{32} - 3 \cdot 10^{33}$~W.  In short, the two techniques give consistent values for the survey sensitivity to continuum/continuum+line luminosity.  The sensitivity to line detections is $\sim \sqrt{2}$ poorer since we derive line strengths via subtraction of two narrowband luminosities, as described in the following section.

\subsection{Source Identification and Flux Extraction}
\label{sec:sextractor}

SExtractor (Bertin \& Arnouts 1996) is used to identify sources and to extract their fluxes.  Extensive simulations were performed to optimize the extraction approach for our imaging program; key parameter settings include {\tt DETECT\_THRESH}=1.5 and {\tt DETECT\_MINAREA}=3 0\farcs523 pixels.
%(detection threshold, minimum number of contiguous pixels above the detection threshold, aperture characteristics, etc.)  
The {\tt FLUX\_AUTO} output parameter (with the default {\tt PHOT\_AUTOPARAMS}=[2.5,3.5]) is used for the flux extractions, as it effectively recovers simulated fluxes for the variety of source morphologies and sizes typical of this survey.  Narrowband filter pairs are employed at each epoch, and it is unknown in advance whether a particular source will produce line emission appearing in `Filter A' or `Filter B' of a given filter pair.  Thus, SExtractor is used in `dual' mode twice for each stack.  First, sources are identified and fluxes are extracted from the Filter A stack, then SExtractor is rerun on the Filter B stack using the same positions and apertures that were determined from Filter A imaging.  The procedure is then repeated in the reverse order: identifications and extractions are based on Filter B imaging, and then the Filter A image is processed using Filter B positions and apertures.  In this way two goals are accomplished: fluxes are consistently extracted from Filter A and Filter B imaging, and line emitters with weak continuum levels are not overlooked.

\begin{figure}
 \plotone{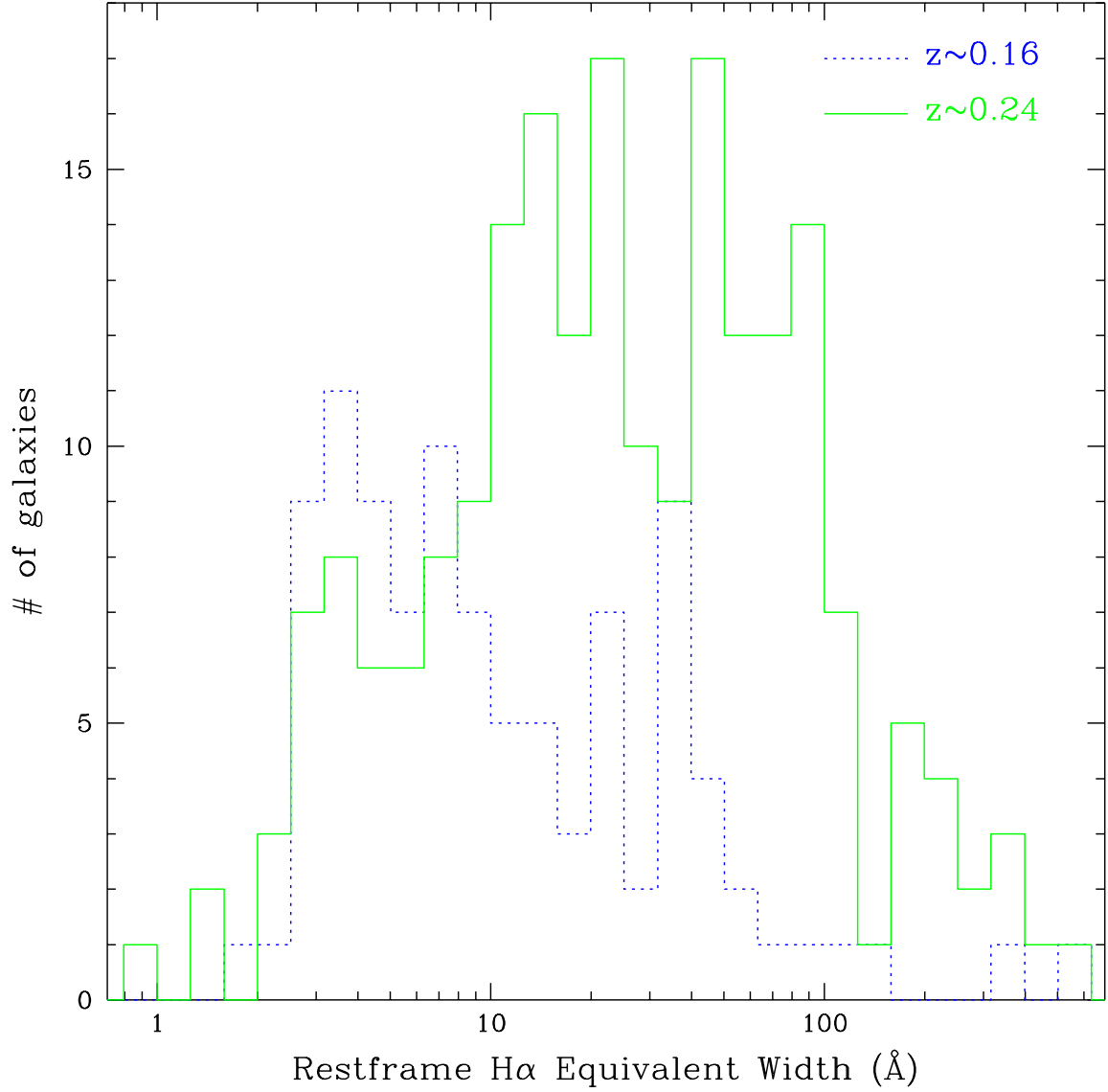}
 \caption{The distribution of \hal\ equivalent widths at $z\sim0.16$ (blue dotted line) and 0.24 (green solid line).}
 \label{fig:ew}
\end{figure}

Traditionally, narrowband imaging surveys rely on broadboand imaging to infer and subtract off the underlying stellar continuum.  Since two wavelength-adjacent but otherwise nearly identical narrowband filters are used at each epoch for this survey, the line emission is derived from a simple subtraction of the flux from Filter A and the flux from Filter B (after accounting for the slightly different filter widths): 
\begin{eqnarray}
f({\rm H}\alpha+[{\rm NII}]) &=& f_{\rm A} - f_{\rm B}{\Delta_{\rm A} \over \Delta_{\rm B}} \;\;\; {\rm if} \;\;\; f_{\rm A} > f_{\rm B}{\Delta_{\rm A} \over \Delta_{\rm B}}\\ 
f({\rm H}\alpha+[{\rm NII}]) &=& f_{\rm B} - f_{\rm A}{\Delta_{\rm B} \over \Delta_{\rm A}} \;\;\; {\rm if} \;\;\; f_{\rm B} > f_{\rm A}{\Delta_{\rm B} \over \Delta_{\rm A}}\\
f({\rm H}\alpha) &=& 0.8 f({\rm H}\alpha+[{\rm NII}]) \label{eq:Ha} \\
L({\rm H}\alpha) &=& 4\pi D_{\rm L}^2 f({\rm H}\alpha) \\
{\rm S/N}({\rm H}\alpha) &=& f({\rm H}\alpha+[{\rm NII}]) / \sqrt{\epsilon(f_A)^2 + \epsilon(f_B)^2} 
\end{eqnarray}
where a correction for \NII\ that is typical of star-forming galaxies is employed (Kennicutt 1992; Jansen et al. 2000).  The above expressions include the luminosity distance $D_{\rm L}$ and filter fluxes derived as the filter width times the flux density, e.g., $f_{\rm A}=\Delta_{\rm A}{f_{\rm A}}_{\lambda}$.  While our dual-narrowband imaging is more accurate than a traditional narrowband+broadband observational approach, there still may be a selection effect working against low equivalent width sources.  Hanish et al. (2006) show that 4.5\% of their \hal\ luminosity density comes from galaxies with equivalent widths smaller than 10\AA.  Figure~\ref{fig:ew} displays the \hal\ equivalent width distributions for the two epochs studied in this work, showing that a significant portion of our sources have equivalent widths below 10\AA.  The distributions are qualitatively in agreement with that presented in the literature (e.g., Gronwall et al. 2004; Haines et al. 2007), though there is evidence for larger equivalent widths at redshift $z\sim0.24$ than at $z\sim0.16$.  The median equivalent widths are 8.5\AA\ and 23\AA\ at redshifts 0.16 and 0.24, respectively.  

The redshift of each line emitter is not known {\it a priori} and thus detections from galaxies outside of the target epochs could bias to high levels the \hal\ detection rate and thus the inferred star formation rate density at each epoch; other prominent optical emission lines such as \OIII $\lambda\lambda$4959/5007, \OII $\lambda$3727, and \hbeta $\lambda$4861 could be redshifted into the filter bandpasses (e.g., Fujita et al. 2003).  Photometric redshifts are used to cull the contaminators.  Multiple photometric redshift catalogs are or soon will be available for the fields pursued in this work: Rowan-Robinson et al. (2007) from SWIRE, Csabai et al. (2003) from SDSS, and Moore et al. (in preparation) from WIRO observations.  The SWIRE survey utilizes a suite of optical and infrared fluxes along with a variety of spiral, elliptical, and AGN SED templates to determine photometric redshifts.  The Sloan survey uses a similar technique with $ugriz$ data, and our WIRO-based catalog employs $UBVRI$ imaging in addition to our narrowband data.  In regions where the coverage overlaps (e.g., Lockman Hole), we have compared the impact of using photometric redshifts from the various surveys.  Survey products such as the parameters and integrals of the \hal\ luminosity functions are fairly insensitive to which survey is utilized for photometric redshifts (Moore et al., in preparation).  However, for the sake of uniformity SWIRE photometric redshifts are used where available, with SDSS and our own photometric redshifts employed to supplement the SWIRE coverage.

The uncertainty on these photometric redshifts is a function of redshift, flux, the number and rest wavelengths of the broadband detections, the SED models, etc.  The typical photometric redshift used in this work has an uncertainty of $\sim$5\%.  A source is considered to be at one of the targeted redshifts if its photometric redshift falls within one of our filter's effective bandpasses, i.e., within $\Delta$/2 of one our filter's central wavelengths (see \S~\ref{sec:observations}).  The sample of \hal\ emitters is created after applying a 3$\sigma$ cut based on the \hal\ signal-to-noise.  The large number of sources detected in our survey leads to a robust {\it statistical} representation of the star formation rate at each cosmic epoch, but the reliance on photometric redshifts implies a certain level of error in tagging {\it individual} sources as residing within the targeted redshift ranges.  The impact of relying on photometric redshifts will be more fully explored in a future contribution that presents a larger fraction of the survey.

\section {Results}
\label{sec:results}

A primary tool for understanding the basic characteristics of a galaxy population is the luminosity function (Schechter 1976).  The \hal\ luminosity function is constructed for each epoch assuming a Schechter profile:
\be
\Phi(z,\log L) d\log L = \phi(z,L) dL = \phi_*(z) \left( {L \over L_*(z)} \right)^{\alpha(z)} e^{-L/L*(z)} {dL \over L_*(z)},
\ee
where $\alpha(z)$ conveys the shape of the function, $L_*(z)$ sets the luminosity scale, and $\phi_*(z)$ represents the overall normalization.  The computation of the $i^{\rm th}$ bin of the luminosity function and its uncertainty follows
\be
\Phi(z,\log L_i) = {\Sigma_j V(z_j,L_j)^{-1} \over \Delta \log L}, \;\;\;\;\;\;\; \epsilon[\Phi(z,\log L_i)] = {\sqrt {\Sigma_j V(z_j,L_j)^{-2}} \over \Delta \log L}
\label{eq:lf}
\ee
where $V(z_j,L_j)$ is the comoving volume for the $j^{\rm th}$ galaxy in the summation, $\mid \log L_j - \log L_i \mid < \onehalf \Delta \log L$, and the bin width $\Delta \log L$ spans 0.4~dex.  The \hal\ luminosity functions for each epoch are presented and discussed after a brief description below of corrections that are first applied to the data.

\subsection {Accounting for Observational Biases}
\label{sec:biases}

\subsubsection{Non-Rectangular Filter Transmission Profiles}
If each filter bandpass were to follow a `tophat' profile, then the comoving volume computed in Equation~\ref{eq:lf} would be simply a product of the filter width $\Delta$ and the number of square degrees surveyed (see Table~\ref{tab:filters}). However, since the filter bandpasses are not tophats, it is more difficult to detect faint galaxies over the full bandpass.  Hence, the volume in Equation~\ref{eq:lf} for faint galaxies is somewhat smaller than $\Delta$.  The volume correction factors, $\zeta(z)_{\rm filter}$, where
\be
V(z,L)_{\rm cor} = \zeta(z)_{\rm filter} V(z,L),
\ee
can be as large as 10-20\%, but the median volume correction factors are $\zeta_{\rm filter}$=0.985, 0.981, 0.990, and 0.988, respectively for filters centered at 7597\AA, 7661\AA, 8132\AA, and 8199\AA.

A second effect is related to the non-rectangular nature of the filter bandpass profiles.  If spectra were available to provide accurate measures of individual galaxy redshifts, then each \hal\ luminosity could be corrected by the inverse of the normalized filter transmission at the appropriate wavelength, $\mathcal{T}(\lambda)^{-1}$.  Since such spectra are not available for this survey, galaxies that are detected in the low-transmission wings of the filter profiles are measured to have artificially faint \hal\ luminosities.  A statistical correction, $\eta_{\rm filter}$, can be applied to all \hal\ luminosities to remedy this effect.  The correction is derived by calculating the average inverse normalized transmission for each filter, weighted by the survey's distribution of detectable \hal\ luminosities (assuming a Schechter luminosity function).  Monte Carlo simulations of the effect indicate $\eta_{\rm filter} \approx 1.10\pm0.08$.

\subsubsection{Extinction}
Fluxes are corrected for foreground Milky Way extinction, an effect on the order of 1\% for these regions of low Galactic cirrus emission.  More importantly, the \hal\ data are also corrected for the internal extinction within the survey galaxies.  A standard recipe is to assume 1 magnitude of extinction at either $V$ band (a factor of $\sim$2.1 at 6563\AA) or at the \hal\ wavelength (a factor of $\sim$2.5 at 6563\AA).  Another approach is to employ an extinction that varies with galaxy luminosity.  Hopkins et al. (2001), for example, suggest a transcendental equation for the extinction based on the galaxy star formation rate:
\be
\log {\rm SFR(H}\alpha)_{\rm obs}=\log {\rm SFR(H}\alpha)_{\rm int}-\alpha \log {\left[ \beta \log{\rm SFR(H}\alpha)_{\rm int} +\gamma \over \delta \right]},
\ee
where $\alpha=2.360$, $\beta=0.797$, $\gamma=3.786$, and $\delta=2.86$ if a Cardelli et al. (1989) reddening curve is assumed (see Ly et al. 2007).  This approach is adopted for the WySH survey.  Table~\ref{tab:corrections} lists the extinction correction factor, $\theta(L)_{\rm ext}$, for the relevant bins in the luminosity function.  The corrected \hal\ luminosity and star formation rate are thus expressed as
\begin{eqnarray}
L({\rm H}\alpha)_{\rm cor} &=& \eta_{\rm filter} \theta(L)_{\rm ext} L({\rm H}\alpha)_{\rm obs} \\
{\rm SFR}({\rm H}\alpha)~[M_\odot~{\rm yr}^{-1}] &=& 7.9\cdot10^{-35} L({\rm H}\alpha)_{\rm cor}~[{\rm W}] \\
\epsilon({\rm SFR}) &=& {\rm SFR} \sqrt{\epsilon(f_A)^2/f_A^2+\epsilon(f_B)^2/f_B^2+0.02^2+0.15^2+0.05^2+0.20^2} 
\end{eqnarray}
where the Kennicutt (1998) calibration for the star formation rate is utilized.  The uncertainty in the star formation rate, $\epsilon$(SFR), includes factors of 2\%, 15\%, 5\%, and 20\% to account for uncertainties in the filter widths, the correction for \NII\ (Equation~\ref{eq:Ha}), the flux calibration, and the correction factors described in this section, respectively.

\subsubsection{Sample Incompleteness}
\label{sec:incompleteness}
An important correction accounts for the increasingly limited ability to detect sources with fainter \hal\ luminosities.  An incompleteness correction, $\kappa(z,L)_{\rm inc}^{-1}$, is derived for each bin of the luminosity function, for each epoch observed,  
\be
\Phi(z,\log L)_{\rm cor} = \kappa(z,L)_{\rm inc}^{-1} \Phi(z,\log L)_{\rm obs}
\ee
Extensive Monte Carlo simulations are employed to quantify the impact of incompleteness.  For each luminosity bin in the luminosity function, IRAF/{\tt mkobjects} is used to create several mock images with the same noise as the actual survey image stacks described in Section~\ref{sec:observations}.  These mock images incorporate typical seeing effects and generate sources with distributions similar to those of actual sources in spatial density, equivalent widths, \hal\ diameters (Dale et al. 1999), and of course the luminosity of the particular luminosity bin being simulated.  Table~\ref{tab:corrections} provides a compilation of incompleteness factors $\kappa(z,L)_{\rm inc}$ for samples that are selected with a 3$\sigma$ cut
based on the \hal\ signal-to-noise.  The uncertainties in the incompleteness are based on the ranges in incompleteness for a variety of simulation parameters (e.g, 500 vs. 1500 vs 4500 sources in a simulated image; $1<$EW$(\AA)<100$ vs $10<$EW$(\AA)<50$, etc).  For brighter luminosities the incompleteness reaches an asymptotic value of $\sim0.95$ due to the effects of sources overlapping one another.  Additional pseudo-empirical simulations, whereby artificial sources are added to actual image stacks, indicate consistent results with the `pure' simulations.

\subsubsection{Active Galaxies and Star-Forming Galaxies Lacking \hal\ Emission}
A main goal of this survey is to derive the volume-averaged star formation rate at several epochs, $\dot{\rho}_{\rm SFR}(z)$.  If $\dot{\rho}_{\rm SFR}(z)$ is to be interpreted as being solely due to star-forming galaxies, then contributions from AGN galaxies will have to be removed.  Brinchmann et al. (2004) and Salim et al. (2007), for example, respectively use multi-wavelength data for $\sim150,000$ and $\sim50,000$ galaxies to derive a $\sim$4\% AGN contribution to $\dot{\rho}_{\rm SFR}(z)$.  However, these same studies estimate between $\sim$1 and 11\% of the cosmic star formation rate stems from galaxies lacking obvious optical line emission.  These two effects have a counteracting impact on $\dot{\rho}_{\rm SFR}(z)$, and in light of their uncertainties and relatively small impact on the final outcome, we do not invoke any corrections to account for them.

\subsection {Preliminary Luminosity Functions at $z\sim0.16$ and 0.24}
\begin{figure}
 \plotone{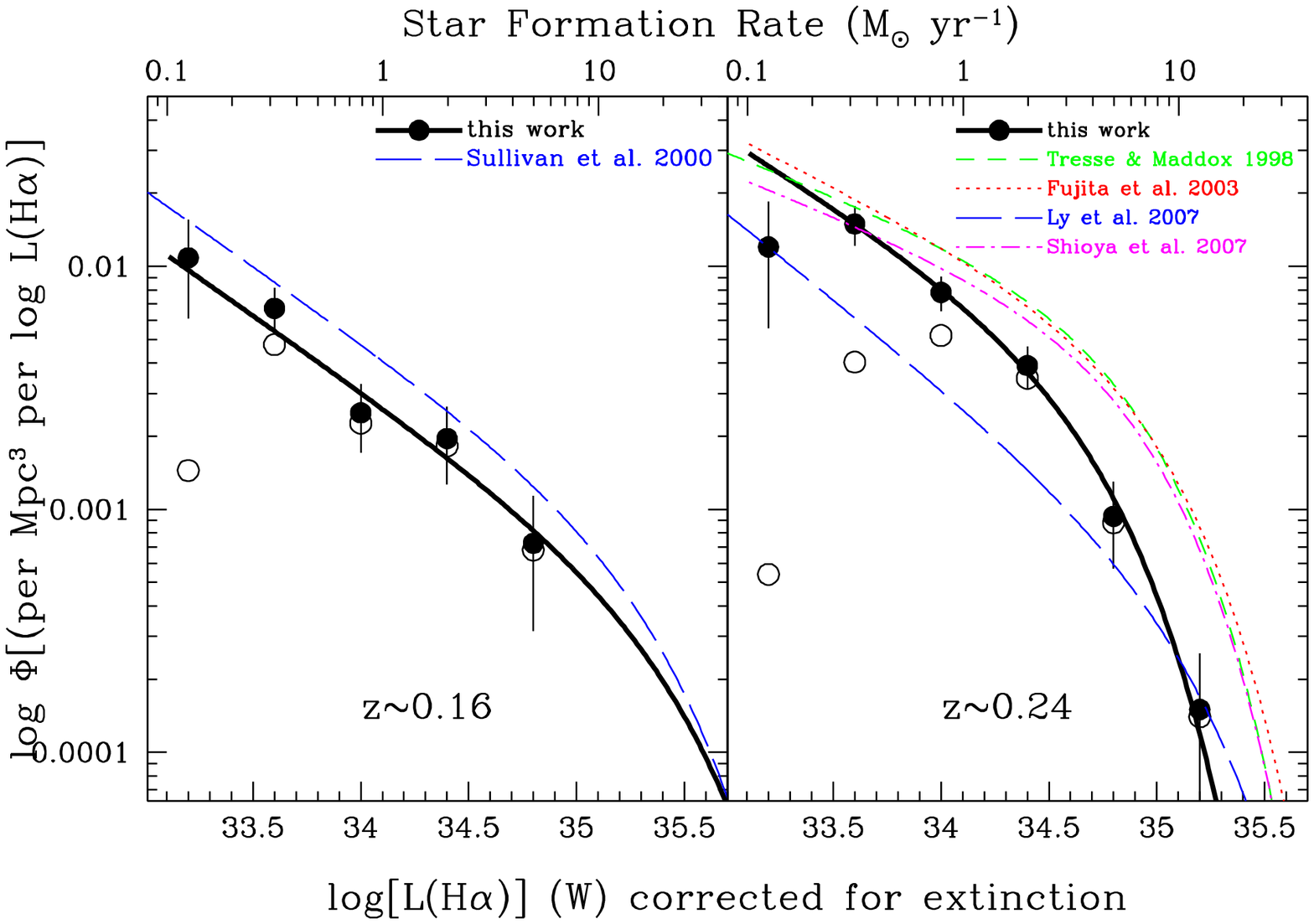}
 \caption{The preliminary luminosity functions at $z\sim$0.16 and 0.24, based on \Nsourcesa +\Nsourcesb\ galaxies observed over $\sim$\sqdeg\ square degrees.  The data without incompleteness corrections (\S~\ref{sec:incompleteness}) are displayed as open circles, while those corrected for incompleteness are shown as filled circles.  Error bars reflect the uncertainty in the luminosity function amplitude according to Equation~\ref{eq:lf}, summed in quadrature with the uncertainties in the incompleteness corrections.  The thick solid lines show the Schechter fits for all luminosity bins except $L({\rm H}\alpha)=10^{33.2}$~W at $z\sim0.24$; the parameters for these fits are presented in the first two rows of Table~\ref{tab:schechter}.}
 \label{fig:lfs}
\end{figure}

Figure~\ref{fig:lfs} shows the initial luminosity functions at $z\sim0.16$ and 0.24 for the WySH survey.  Open circles indicate the data corrected for all issues described above except incompleteness; the filled circles also include corrections for incompleteness.  The thick solid lines show the Schechter fits for all luminosity bins except $L({\rm H}\alpha)=10^{33.2}$~W at $z\sim0.24$.  The parameters for the displayed fits are listed in the first two rows of Table~\ref{tab:schechter}; for comparison with the literature (e.g., Table~5 of Ly et al. (2007)), we also include in Table~\ref{tab:schechter} fit parameters assuming 1 and 0~mags of internal extinction at the \hal\ wavelength.  
%Note that the slope $\alpha$ is generally fixed due to the small number of bins satisfying $L({\rm H}\alpha)\geq10^{34}$~W, though we do provide one set of fits in Table~\ref{tab:schechter} where the slope is treated as a free parameter (with relatively large uncertainty).  
Tentatively, we observe a significantly higher luminosity function amplitude $\phi_*$ at $z\sim0.24$ than at $z\sim0.16$.
It is difficult to compare the characteristic luminosity $L_*$ between the two epochs since it is poorly defined at $z\sim0.16$.
%, and a significantly brighter characteristic luminosity $L_*$ at $z\sim0.24$ (a factor of 2 for the fits displayed in Figure~\ref{fig:lfs}).  
Error bars in Figure~\ref{fig:lfs} reflect the uncertainty in the luminosity function amplitude according to Equation~\ref{eq:lf}, summed in quadrature with the uncertainties in the incompleteness corrections.  Also included in Figure~\ref{fig:lfs} are \hal-based luminosity functions from the literature, largely consistent with our preliminary results.

\subsection {The Cosmic Star Formation Rate Density at $z\sim0.16$ and 0.24}

The volume-averaged cosmic star formation rate can be computed by integrating under the fitted Schechter function and multiplying by the Kennicutt (1998) star formation rate calibration:
\be
\dot{\rho}_{\rm SFR} (h_{70} M_\odot {\rm yr}^{-1} {\rm Mpc}^{-3})= 7.9\cdot10^{-35} \mathcal{L}({\rm W})
\ee
where an analytical expression for the luminosity density is
\be
\mathcal{L} = \int_0^{\infty} dL L \Phi(L) = \phi_{\star}L_{\star}\Gamma(\alpha+2).
\ee
The larger $\phi_*$ and brighter $L_*$ at $z\sim0.24$ lead to an overall larger cosmic star formation rate density at that redshift, by a factor of 1.5 for the canonical fits displayed in Figure~\ref{fig:lfs}.  Table~\ref{tab:schechter} provides the integrated cosmic star formation rate densities.  For comparison with other extinction-corrected, \hal-based cosmic star formation rate densities, we are comfortably within a factor of 2 of the values published by Sullivan et al. (2000) at $z \sim 0.16$ and Pascual et al. (2001), Fujita et al. (2003), Hippelein et al. (2003), Ly et al. (2007), Westra \& Jones (2007), and Shioya et al. (2007) at $z \sim 0.24$ (see Ly et al. 2007 for values converted to the common cosmology also assumed in this work).

The formal uncertainties presented in Table~\ref{tab:schechter} are based solely on the fitting procedure, weighted by the uncertainties in the bin data points as described by Equation~\ref{eq:lf}.  Also to be considered is the impact of ``cosmic variance'', fluctuations due to the characteristics of the particular volume(s) being probed along a survey's line(s)-of-sight (e.g., clusters, voids, etc.).  An initial estimate of the WySH cosmic variance can be made by performing a separate analysis of the two fields studied in this presentation (Lockman Hole and ELAIS-N1).  Preliminarily, the cosmic star formation rate densities in these two fields differ by $\sim$25\% at each epoch, indicating a modest impact due to cosmic variance.
%Uncertainties in the LFs due to cosmic variance are described in Davis \& Huchra (1982) for estimating cosmic variance for a given volume; this is what Huang et al. astro-ph/0704.3609 say.

\section {Summary}
\label{sec:summary}
We report first results from the two most recent epochs of WySH, the {\it Wyoming Survey for H$\alpha$}.  The overall scientific goal of the survey is to accurately calibrate changes in the \hal\ luminosity function via $\lesssim$4 square degrees of narrowband imaging from several epochs spanning the latter half of the Universe.  From a technical standpoint, our aim is to achieve relatively small statistical uncertainties by executing a luminosity-uniform survey for hundreds of galaxies at each epoch using narrowband filter pairs for improved continuum subtraction.  Preliminary results from approximately 2 square degrees from two fields indicate a clear evolution in the volume-averaged cosmic star formation rate, a decrease by a factor of $\sim$1.5 from a redshift of $z\sim0.24$ to a redshift of $\sim$0.16.  Our value at $z\sim0.16$ is comparable to that found from \hal\ work at $z=0$ by Gallego et al. (1995) and Hanish et al. (2006), so more substantial evolution appears to have occurred between redshifts of 0.16 and 0.24 compared to what has occurred after a redshift of 0.16, but additional data are needed to more robustly assess this claim.  By separately analyzing these initial data from the Lockman Hole and ELAIS-N1 fields, we see a $\sim$25\% ``cosmic variation'' in our cosmic star formation rate densities at $z\sim0.16$ and 0.24.  The estimate of the cosmic variance in our survey will be more accurately estimated in future work, after more data are added from ELAIS-N1 and Lockman, and data are included from our third field (``WySH~1'').  
%The WySH team is the best ever!

\acknowledgements 
This research is funded through the NSF CAREER and REU programs (AST0348990 and AST0353760) and the Wyoming NASA Space Grant Consortium (NNG05G165H).  This survey would not have occurred without the enterprise of Michael Pierce in developing WIROPrime, the assistance of Chip Kobulnicky, Andy Monson, and Steve Hodder in fabricating the prime focus corrector lens, and the diligence of James Weger and Josh Silvey in keeping WIRO operational.  
%We acknowledge contributions from many affiliated members of the WySH survey: Christine Lamanna, Travis Laurance, Mark Reiser, Shianne Kattner, Janice Lee, Rose Finn, and the many students and science teachers who helped with the observations. 
We acknowledge contributions from WySH affiliates Travis Laurance, Mark Reiser, Janice Lee, Rose Finn, and the many students and science teachers who helped with the observations.  We appreciate the early access to the photometric redshifts provided by the SWIRE team, and fruitful discussions with Chun Ly.
%This research has made use of the NASA/IPAC Extragalactic Database which is operated by JPL/Caltech, under contract with NASA.  
This research has made use of the NASA/IPAC Infrared Science Archive, which is operated by the Jet Propulsion Laboratory, California Institute of Technology, under contract with NASA.  IRAF, the Image Reduction and Analysis Facility, has been developed by the National Optical Astronomy Observatories and the Space Telescope Science Institute.  The Digitized Sky Surveys were produced at STScI (NAG~W-2166).  The images of these surveys are based on photographic data obtained using the Oschin Schmidt Telescope on Palomar Mountain and the UK Schmidt Telescope.  Funding for the Sloan Digital Sky Survey and SDSS-II has been provided by the Alfred P. Sloan Foundation, the Participating Institutions, the NSF, the U.S. Department of Energy, NASA, the Japanese Monbukagakusho, the Max Planck Society, and the Higher Education Funding Council for England.

%%%%%%%%%%%%%%%%%%%%%%%%%%%%%%%%%%%%%%%%%%%%%%%%%%%%%%
%TABLES
%%%%%%%%%%%%%%%%%%%%%%%%%%%%%%%%%%%%%%%%%%%%%%%%%%%%%%
\begin{deluxetable}{lccc}
\tabletypesize{\scriptsize}
\tablenum{1}
\label{tab:fields}
\tablecaption{Target Fields}
\tablewidth{0pc}
\tablehead{
\colhead{Name} &
\colhead{R.A.~\&~Dec.} &
\colhead{$I_\nu(100\mu$m)} &
\colhead{Observing} 
\\
\colhead{} &
\colhead{(J2000)} &
\colhead{(MJy~sr$^{-1}$)} &
\colhead{Window} 
}
\startdata
WySH 1      &00~12~40~~$+$32~49~00&$\sim$1.0&Jul-Nov\\
Lockman Hole&10~47~10~~$+$58~23~00&$\sim$0.9&Nov-Jun\\
ELAIS N1    &16~11~10~~$+$55~22~00&$\sim$0.4&Feb-Sep\\
\enddata
\tablecomments{Units of right ascension are hours, minutes, and seconds, and units of declination are degrees, arcminutes, and arcseconds.}
\end{deluxetable}
%%%%%%%%%%%%%%%%%%%%%%%%%%%%%%%%%%%%%%%%%%%%%%%%%%%%%%%%%%%%%%%%%%%%%%%%%%%%%%%%%%%
\begin{deluxetable}{ccccccccc}
\tabletypesize{\scriptsize}
\tablenum{2}
\label{tab:filters}
\tablecaption{Optical Filters and Survey Parameters}
\tablewidth{0pc}
\def\a{\tablenotemark{a}}
\def\b{\tablenotemark{b}}
\def\c{\tablenotemark{c}}
\tablehead{
\colhead{Filter}     &
\colhead{Filter}     &
\colhead{Redshift}   &
\colhead{Luminosity} &
\colhead{Current}    &
\colhead{Comoving}   &
\colhead{Integration}&
\colhead{Sensitivity\c}&
\colhead{Sensitivity\c}       
\\
\colhead{${\bar \lambda}$} &
\colhead{$\Delta$}         &
\colhead{for \hal}         &
\colhead{Distance}         &
\colhead{Area\a}           &
\colhead{Volume/Area}      &
\colhead{per pixel}        &
\colhead{(3$\sigma$)}      &
\colhead{(3$\sigma$)}       
\\
\colhead{($\AA$)}                 &
\colhead{($\AA$)}                 &
\colhead{}                        &
\colhead{($h^{-1}_{70}$ Mpc)}     &
\colhead{($\sq\degr$)}               &
\colhead{($h^{-3}_{70}$ Mpc$^3/\sq\degr$)} &
\colhead{(sec)}                   &
\colhead{(10$^{-20}$W/m$^2$)}            &       
\colhead{(10$^{32}$W)}                  
}
\startdata
7597&60.97&0.1575&~753&2.18   &~4745&1200&6.3  &4.3  \\ %&~5557&16610
7661&60.54&0.1673&~804&2.18   &~5263&1200&6.0  &4.7  \\ %&~6164&18420
8132&57.68&0.2392&1198&1.92   &~9512&3600&4.1  &7.1  \\ %&10960&33290
8199&56.55&0.2493&1256&1.92   &10030&3600&3.2  &6.1  \\ %&11550&35090
8614&58.56&0.3126&1628&\nodata&15260&6000&2.8\b&8.7\b\\ %&n/a  &53390
8687&64.10&0.3237&1695&\nodata&17690&6000&2.8\b&9.5\b\\ %&n/a  &61920
9155&57.66&0.3950&2140&\nodata&21920&9600&2.2\b&12\b \\ %&n/a  &76720
9233&58.75&0.4068&2216&\nodata&23390&9600&2.2\b&13\b \\ %&n/a  &81850
\enddata
\tablenotetext{a}{\footnotesize The projected ultimate coverage for the entire optical survey is 3.5-4 square degrees.}
\tablenotetext{b}{\footnotesize Planned.}
\tablenotetext{c}{\footnotesize For a 3\arcsec\ diameter aperture, as explain in \S~\ref{sec:sensitivity}.}
\end{deluxetable}
%%%%%%%%%%%%%%%%%%%%%%%%%%%%%%%%%%%%%%%%%%%%%%%%%%%%%%%%%%%%%%%%%%%%%%%%%%%%%%%%%%%
\begin{deluxetable}{cccc}
\tabletypesize{\scriptsize}
\tablenum{3}
\label{tab:corrections}
\tablecaption{H$\alpha$ Extinction and Luminosity Function Incompleteness}
\tablewidth{0pc}
\def\p{$\pm$}
\tablehead{
\colhead{log$L$(H$\alpha$)} &
\colhead{$\theta(L)_{\rm ext}$} &
\colhead{$\kappa(z,L)_{\rm inc}$}      &
\colhead{$\kappa(z,L)_{\rm inc}$}  
\\         
\colhead{(W)} &
\colhead{} &
\colhead{7597/7661\AA}      &
\colhead{8132/8199\AA}           
}
\startdata
33.2&1.18&0.13\p0.02&0.05\p0.01\\
33.6&1.30&0.71\p0.05&0.27\p0.04\\
34.0&1.84&0.91\p0.05&0.66\p0.07\\
34.4&2.38&0.93\p0.04&0.89\p0.05\\
34.8&2.68&0.94\p0.04&0.94\p0.04\\
35.2&3.17&0.95\p0.03&0.95\p0.03\\
\enddata
\end{deluxetable}
%%%%%%%%%%%%%%%%%%%%%%%%%%%%%%%%%%%%%%%%%%%%%%%%%%%%%%%%%%%%%%%%%%%%%%%%%%%%%%%%%%%
\begin{deluxetable}{cllccc}
\tabletypesize{\scriptsize}
\tablenum{4}
\label{tab:schechter}
\tablecaption{Luminosity Function Results}
\tablewidth{0pc}
\def\a{\tablenotemark{a}}
\def\b{\tablenotemark{b}}
\def\c{\tablenotemark{c}}
\def\p{$\pm$}
\tablehead{
\colhead{Redshift}  &
\colhead{Extinction}&
\colhead{$\alpha$}&
\colhead{$\log L_*$}&
\colhead{$\log \phi_*$}&
\colhead{$\dot{\rho}_{\rm SFR}$\c}
\\
\colhead{}    &
\colhead{Factor}&
\colhead{}    &
\colhead{(W)} &
\colhead{($h_{70}^3 {\rm Mpc}^{-3}$)}  &
\colhead{$h_{70} M_\odot {\rm yr}^{-1} {\rm Mpc}^{-3}$}
}
\startdata
%0.16&1.18-3.17\a &-1.62\p0.05\b&34.9\p0.2&-3.22\p0.20&0.009\p0.006\\ % forced slope
%0.24&1.18-3.17\a &-1.50\p0.05\b&35.2\p0.1&-3.03\p0.10&0.019\p0.007\\ % forced slope
0.16&1.18-3.17\a &-1.62\p0.05  &35.5\p1.2&-3.81\p0.81&0.009\p0.006\\
0.24&1.18-3.17\a &-1.55\p0.15  &34.7\p0.2&-2.79\p0.23&0.014\p0.007\\
0.16&2.51 (1 mag)&-1.62\p0.05\b&34.7\p0.2&-2.94\p0.19&0.011\p0.007\\
0.24&2.51 (1 mag)&-1.55\p0.05\b&34.8\p0.1&-2.71\p0.12&0.019\p0.008\\
0.16&1.00 (0 mag)&-1.62\p0.05\b&34.1\p0.1&-2.66\p0.13&0.006\p0.002\\
0.24&1.00 (0 mag)&-1.55\p0.05\b&35.0\p0.1&-2.94\p0.08&0.019\p0.007\\
\enddata
%\tablenotetext{a}{\footnotesize Forced to match the slope at $z\sim0.24$.}
\tablenotetext{a}{\footnotesize See Table~\ref{tab:corrections}.}
\tablenotetext{b}{\footnotesize Fixed; not fitted.}
\tablenotetext{c}{\footnotesize The quoted uncertainties are statistical; cosmic variance is tentatively estimated at 25\%.}
\end{deluxetable}
%%%%%%%%%%%%%%%%%%%%%%%%%%%%%%%%%%%%%%%%%%%%%%%%%%%%%%%%%%%%%%%%%%%%%%%%%%%%%%%%%%%
 
%%%%%%%%%%%%%%%%%%%%%%%%%%%%%%%%%%%%%%
%FIGURES
%%%%%%%%%%%%%%%%%%%%%%%%%%%%%%%%%%%%%%

\begin{figure}
% \plotone{f1.eps}
 \caption{The $\sim$18\arcmin\ fields targeted at WIRO in the WySH~1 region are shown as squares overlaid on an IRAS 100\m\ image.}
 \label{fig:wysh1}
\end{figure}

\begin{figure}
% \plotone{f2.eps}
 \caption{The $\sim$18\arcmin\ fields targeted at WIRO in the Lockman Hole region are shown as small squares overlaid on an IRAS 100\m\ image.  The 9.3 square degrees mapped at 3.5, 4.5, 5.8, and 8.0\m\ by the SWIRE program is outlined in bold, while the 9.2 square degrees mapped at 24, 70, and 160\m\ by SWIRE is outlined with a thinner line.  Fields from the GALEX Deep Imaging Survey are shown as circles with $\sim$0\fdg6 radii.} 
 \label{fig:lockman}
\end{figure}

\begin{figure}
% \plotone{f3.eps}
 \caption{The $\sim$18\arcmin\ fields targeted at WIRO in the ELAIS~N1 region are shown as small squares overlaid on an IRAS 100\m\ image.  The 11.1 square degrees mapped at 3.5, 4.5, 5.8, and 8.0\m\ by the SWIRE program is outlined in bold, while the 11.0 square degrees mapped at 24, 70, and 160\m\ by SWIRE is outlined with a thinner line.  Fields from the GALEX Deep Imaging Survey are shown as circles with $\sim$0\fdg6 radii.} 
 \label{fig:elaisn1}
\end{figure}

\clearpage
\begin {thebibliography}{dum}
\bibitem[]{}Ascasibar, Y., Yepes, G., Gottlober, S. \& Muller, V. 2002, \aap, 387, 396
\bibitem[]{}Babbedge, T.S.R. et al. 2006, \mnras, 370, 1159
\bibitem[]{}Baldry, I.K. et al. 2002, \apj, 569, 582
%\bibitem[]{}Balogh, M.L., Morris, S.L., Yee, H.K.C., Carlberg, R.G. \& Ellingson, E. 1997, \apj, 488, 75
\bibitem[]{}Balogh, M.L., Couch, W.J., Smail, I., Bower, R.G. \& Glazebrook, K. 2002, \mnras, 335, 10
%\bibitem[]{}Balogh, M.L., Bower, R.G., Smail, I., Ziegler, B.L., Davies, R.L., Gaztelu, A. \& Fritz, A. 2002a, \mnras, 337, 256
%\bibitem[]{}Balogh, M.L., Smail, I., Bower, R.G., Ziegler, B.L., Smith, G.P., Davies, R.L., Gaztelu, A., Kneib, J.-P. \& Ebeling, H. 2002b, \apj, 566, 123
\bibitem[]{}Bertin, E. \& Arnouts, S. 1996, \aaps, 117, 393
\bibitem[]{}Blain, A.W., Smail, I., Ivison, R.J. \& Kneib, J.-P., 1999, \mnras, 302, 632
\bibitem[]{}Bohlin, R.C., Dickinson, M.,E., \& Calzetti, D. 2001, \aj, 122, 2118
%\bibitem[]{}Boyle, B.J. \& Terlevich, R.J. 1998, \mnras, 293, L49
\bibitem[]{}Brinchmann, J., Charlot, S., White, S.D.M., Tremonti, C., Kauffmann, G., Heckman, T., \& Brinkmann, J. 2004, \mnras, 351, 1151
\bibitem[]{}Burgarella, D., Le Floc'h, E., Takeuchi, T.T., Huang, J.S., Buat, V., Rieke, G.H., \& Tyler, K.D. 2007, \mnras, in press
\bibitem[]{}Calzetti, D. et al. 2007, \apj, in press
\bibitem[]{}Cardelli, J.A., Clayton, G.C., \& Mathis, J.S. 1989, \apj, 345, 245
\bibitem[]{}Cowie, L.L., Songaila, A. \& Barger, A.J. 1999, \aj, 118, 603
\bibitem[]{}Csabai, I. et al. 2003, \aj, 125, 580
\bibitem[]{}Dale, D.A., Giovanelli, R., Haynes, M.P., Campusano, L. \& Hardy, E. 1999, \aj, 118 1468
%\bibitem[]{}Dale, D.A. \& Uson, J. 2000, \aj, 120, 552
\bibitem[]{}Dale, D.A., Giovanelli, R., Haynes, M.P., Hardy, E. \& Campusano, L. 2001, \aj, 121, 1886
%\bibitem[]{}Dale, D.A., Helou, G., A. Contursi, N. Silbermann \& S. Kolhatkar 2001, \apj, 549, 215
%\bibitem[]{}Dale, D.A. \& Helou, G. 2002, \apj, 576, 159
%\bibitem[]{}Dale, D.A. \& Uson, J. 2003, \aj, 126, in press (astro-ph/0304371)
%\bibitem[]{}Drinkwater, M.J. et al. 2001, \mnras, 326, 1076
%\bibitem[]{}Dwek, E. \& Barker, M.K. 2002, \apj, 2002, 575, 7
\bibitem[]{}Davis, M., \& Huchra, J. 1982, \apj, 254, 437
\bibitem[]{}Dole, H. et al. 2006, \aap, 451, 417 
\bibitem[]{}Finn, R.A., Zaritsky, D., McCarthy, D.W., Poggianti, B., Rudnick, G., Halliday, C., Milvang-Jensen, B., Pell\'o, R., \& Simard, L. 2005, \apj, 630, 206
\bibitem[]{}Fujita, S.S, Ajiki, M., Shioya, Y. et al. 2003, \apjl, 586, L115
%\bibitem[]{}Gal, R.R., de Carvalho, R.R, Odewahn, S.C., Djorgovski, S.G. \& Margoniner, V.E. 2000, \aj, 119, 12
\bibitem[]{}Gabasch, A. et al. 2004, \apjl, 616, 83
\bibitem[]{}Gallego, J., Zamorano, J., Aragon-Salamanca, A. \& Rego, M. 1995, \apjl, 455, L1 
\bibitem[]{}Glazebrook, K., Baldry, I.K., Blanton, M.R. et al. 2003, \apj, 587, 55
\bibitem[]{}Gronwall, C. 1999, in {\it After the Dark Ages: When Galaxies were Young (the Universe at $2 < z < 5$)}, eds. S. Holt and E. Smith, (American Institute of Physics Press), p. 335
\bibitem[]{}Gronwall, C., Salzer, J.J., Sarajedini, V.L., Jangren, A., Chomiuk, L., Moody, J.W., Frattare, L.M., \& Boroson, T.A. 2004, \aj, 127, 1943
%\bibitem[]{}Guzzo, L. 2002, in {\it DARK 2002, 4th Heidelberg Int. Conference on Dark Matter in Astro- and Particle Physics}, eds. H.-V. Klapdor-Kleingrothaus \& R. Viollier (Springer); astro-ph/0207285
\bibitem[]{}Hanish, D.J. et al. 2006, \apj, 649, 150
%\bibitem[]{}Haynes, M.P., Giovanelli, R. \& Chincarini, G. 1984, \araa, 22, 445
\bibitem[]{}Heavens, A., Panter, B., Jimenez, R., \& Dunlop, J. 2004, Nature, 428, 625
\bibitem[]{}Hippelein, H., Maier, C., Meisenheimer, K. et al. 2003, \aap, 402, 65
%\bibitem[]{}Hogg, D.W. 2000, astro-ph/9905116
\bibitem[]{}Hopkins, A.M., Connolly, A.J., Haarsma, D.B., \& Cram, L.E. 2001, \aj, 122, 288
\bibitem[]{}Huang, J.-S. et al. 2007, \apj, 664, 840
%\bibitem[]{}Iglesias-Paramo, J. \& Vilchez, J.M. 2001, \apj, 550, 204
%\bibitem[]{}Iglesias-Paramo, J., Boselli, A., Cortese, L., Vilchez, J.M. \& Gavazzi, G. 2002, \aap, 384, 383
\bibitem[]{}Iglesias-Paramo, et al. 2007, \apj, in press
%\bibitem[]{}Janknecht, E., Baade, R. \& Reimers, D. 2002, \aap, 391, L11
\bibitem[]{}Jansen, R.A., Fabricant, D., Franx, M., \& Caldwell, N. 2000, \apjs, 126, 331
\bibitem[]{}Kennicutt, R.C. 1992, \apj, 388, 310
\bibitem[]{}Kennicutt, R.C. 1998, \apj, 498, 541
\bibitem[]{}Kennicutt, R.C. et al. 2007, \apj, in press
\bibitem[]{}Le Floc'h et al. 2005, \apj, 632, 169
%\bibitem[]{}Lewis, I., Balogh, M. et al. 2002, \mnras, 334, 673
\bibitem[]{}Lilly, S.J., LeFevre, O., Hammer, F. \& Crampton, D. 1996, \apjl, 460, L1
\bibitem[]{}Lonsdale, C.J., et al. 2003, \pasp, 115, 897
\bibitem[]{}Ly, C. et al. 2007, \apj, 657, 738
\bibitem[]{}Madau, P., Ferguson, H.C., Dickinson, M.E., Giavalisco, M., Steidel, C.C. \& Fruchter, A. 1996, \mnras, 283, 1388
\bibitem[]{}Mann, R.G. et al. 2002, \mnras, 332, 549
\bibitem[]{}Martin, D.C. et al. 2005, \apj, 521, 64
\bibitem[]{}Oliver, S., Rowan-Robinson, M. et al. 2000, \mnras, 316, 749
\bibitem[]{}Pascual, S., Gallego, J. Aragon-Salamanca, A. \& Zamorano, J. 2001, \aap, 379, 798
\bibitem[]{}Pascual, S., Gallego, \& Zamorano, J. 2007, \pasp, 119, 30
\bibitem[]{}P\'erez-Gonz\'alez, P.G. et al. 2005, \apj, 630, 82
%\bibitem[]{}Phillips, S., Drinkwater, M.J., Gregg, M.D. \& Jones, J.B. 2001. \apj, 560, 201
\bibitem[]{}Pierce, M.J. \& Nations, H.L. 2002, \baas, 200, 6406 
%\bibitem[]{}Puget, J.-L., Abergel, A., Bernard, J.-P., Boulanger, F., Burton, W.B., Desert, F.-X. \& Hartmann, D. 1996, \aap, 308, L5
%\bibitem[]{}Quilis, V., Moore, B. \& Bower, R. 2000, Science, 288, 1617
\bibitem[]{}Reddy, N.A., Steidel, C.C., Fadda, D., Yan, L., Pettini, M., Shapley, A.E., Erb, D.K., \& Adelberger, K.L. 2006, \apj, 644, 792
\bibitem[]{}Rowan-Robinson, M. 2001, \apj, 549, 745
%\bibitem[]{}Rowan-Robinson, M. et al. 2007, in preparation
%\bibitem[]{}Sakai, S. et al. 2002, \apj, 578, 842
\bibitem[]{}Salim, S., et al. 2007, \apjs, in press
\bibitem[]{}Sawicki, M. \& Thompson D. 2006, \apj, 648, 299
%\bibitem[]{}Schaerer, D. 1999, in {\it Building the Galaxies: from the Primordial Universe to the Present}, ed. F. Hammer et al., Editions Frontieres (Gif-sur-Yvette)
\bibitem[]{}Schechter, P. 1976, \apj, 203, 297
\bibitem[]{}Shioya, Y. et al., 2007, \apj, in press
\bibitem[]{}Somerville, R.S., Primack, J.R. \& Faber, S.M. 2001, \mnras, 320, 504
\bibitem[]{}Steidel, C.C., Giavalisco, M., Pettini, M., Dickinson, M. \& Adelberger, K.L. 1996, \apjl, 462, L17
\bibitem[]{}Thompson, R.I., Eisenstein, D., Fan, X., Dickinson, M., Illingworth, G.,  Kennicutt, R.C. 2006, \apj, 647, 787
\bibitem[]{}Tresse, L. \& Maddox, S.J. 1998, \apj, 495, 691
\bibitem[]{}Tresse, L., Maddox, S.J., Le Fevre, O. \& Cuby, J.G. 2002, \mnras, 337, 369
\bibitem[]{}Westra, E. \& Jones, D.H. 2007, \mnras, in press
\bibitem[]{}Wilson, G., Cowie, L.L., Barger, A.J. \& Burke, D.J. 2002, \aj, 124, 1258
%\bibitem[]{}Xu, C., Lonsdale, C.J., Shupe, D.L., O'Linger, J. \& Masci, F. 2001, \apj, 562, 179
\end {thebibliography}
\end{document}